\documentclass[aps,showpacs,graphicx]{revtex4} 
\usepackage{amssymb} 
\usepackage[dvips]{graphicx} 
\usepackage[english]{babel} 
\usepackage{indentfirst} 
\usepackage{amsxtra} 
\usepackage{amsmath} 
\usepackage{supertabular} 
\usepackage{multirow} 
\usepackage [mathcal]{eucal}
\usepackage {float}  
\usepackage{lscape} 
\usepackage{rotating}
\newcommand{\beq}{\begin{equation}} 
\newcommand{\eeq}{\end{equation}} 
\newcommand{\beqn}{\begin{eqnarray}} 
\newcommand{\eeqn}{\end{eqnarray}} 
\usepackage[usenames]{color} 
\usepackage{ulem}

\begin{document} 
 
\noindent 
\title{Shell closure at $N=34$ and the $^{48}$Si nucleus}
 
\author{ G. Co'$^{\,1,2}$, M. Anguiano$^{\,3}$, A. M. Lallena$^{\,3}$}
\affiliation{
$^1$ Dipartimento di Matematica e Fisica ``E. De Giorgi'', 
  Universit\`a del Salento, I-73100 Lecce, Italy \\ 
$^2$ INFN Sezione di Lecce, Via Arnesano, I-73100 Lecce, Italy \\ 
$^3$ Departamento de F\'\i sica At\'omica, Molecular y 
  Nuclear, Universidad de Granada, E-18071 Granada, Spain \\
}  

\date{\today}

\bigskip 
 
\begin{abstract} 
By using a non-relativistic independent particle model we investigate
the mechanism promoting 34 as new magic number. We carried out Hartree-Fock
plus Bardeen-Cooper-Schrieffer and Quasi-particle Random Phase Approximation
calculations by consistently using the same finite-range interaction in all the three
steps of our approach. We used four Gogny-like interactions, with and without tensor
terms. We find that the shell closure for $N=34$ neutrons appears in isotones with 
$Z<26\,$ protons. The smaller is the proton number, the more evident 
is the shell closure at $N=34$. 
An ideal nucleus to investigate this effect should be $^{48}$Si, as it has been recently
suggested. However, some discrepancies occur between the results obtained 
with the four effective interactions we used concerning 
the position of the two-neutron drip line and, therefore, the existence of $^{48}$Si. 
The experimental identification of this nucleus could shed light about the shell evolution in nuclei 
far from the stability valley and put stringent tests on nuclear structure theories. 
\end{abstract} 

\maketitle 

The existence of magic numbers is the main phenome\-nological
evidence justifying the application of an independent particle
model (IPM) to describe atomic nuclei. 
In this approach, the many-body states are 
Slater determinants of single particle (s.p.) 
states. The s.p. properties, such as energy or angular momentum, 
are meaningful in this model, even though the measured quantities are
only those of the global nuclear system.  

The nuclear ground state in the IPM model is built by 
considering that all the s.p. levels below the Fermi energy 
are occupied in accordance with the Pauli exclusion principle.
In spherical systems, a s.p. state with  angular 
momentum $j$ presents a $2j+1$ degeneracy. The magic 
numbers appear when the occupancy of 
all the s.p. levels below the Fermi energy is at its maximum.
The energy ordering of the s.p. states selects the values 
of the magic numbers, which in the surroundings 
of the stability valley show the well known sequence:
2, 8, 20, 28, 40, 50, 82, 126.

The observation that these magic
numbers are not the same in all the regions of the nuclear chart, 
but change when going far away from the stability 
valley  has been a surprise (see for example the reviews of Gade and Glasmacher \cite{gad08a} and Sorlin and Porquet \cite{sor08} for a survey). 
One of the predicted new magic numbers is 34, whose occurrence 
is due to a subtle interplay between spin-orbit and tensor terms 
of the nuclear interaction \cite{gra14}.  

Several authors have investigated, both theoretically and experimentally, 
the appearance of the $N=34$ closure in Ca and Ti isotopes 
\cite{jan02,lid04a,for04,for05a,hon05,wie13,ste13,lid04b}. 
However, while this shell closure
seems to be well established in $^{54}$Ca, 
no clear conclusions have been draw 
in the case of $^{56}$Ti.

Recently, it has been argued that a more clear signature 
of the emergence of this new magic number would be 
provided by the existence of the neutron rich nucleus $^{48}$Si 
\cite{li18}. This prediction has been formulated by using relativistic IPM calculations.

The magic number 34 appears in the $pf$-shell if the energy  
of the $2p_{1/2}$ s.p. state is smaller than that of the
 $1f_{5/2}$ level and the energy gap between them is significant. 
 In this manner, the occupation of the former state
and the emptiness of the latter one build up a shell closure at 34.

\begin{figure}[!t]
\centerline{\includegraphics[width=2.5in]{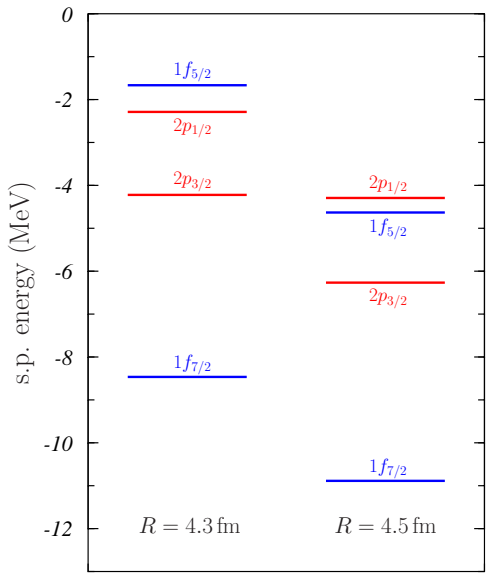}}
\vspace*{8pt}
\caption{Energies of the s.p. levels $2p$ and $1f$ obtained with the Woods-Saxon potential, 
defined in eq. (\ref{eq:WS}). Results for two values of $R$ are shown.\protect\label{fig:ws}}
\end{figure}

The occurrence of this situation 
relies on the subtle combination of different effects.
Let us assume, for example,
that the mean field potential where the nucleons move 
independently of each other is a Woods-Saxon well \cite{boh69}:
\begin{equation}
U(r) \,=\, -V_0\,f(r) \, + \, V_{\rm ls} \,  \frac{r_0^2 }{r} \, \frac{{\rm d}f(r)}{{\rm d} r} \,  {\bf l}\cdot {\bf s}  
\, ,
\label{eq:WS}
\end{equation}
where
\begin{equation}
f(r) \,=\,\frac{1}{1 +\exp\left[ \left( r -R \right) / a \right]}
\end{equation}
and ${\bf l}$ and ${\bf s}$ are, respectively, the s.p. orbital angular
 momentum and spin operators. 
We consider the case with $V_0 = 50\,$MeV, $V_{\rm ls} = 9\,$MeV,  $r_0=1.25\,$fm, $a=0.53\,$fm and $R=4.3\,$fm.
The energies of the two levels of interest
are those shown in the left column of fig. \ref{fig:ws}, and in this case we obtain the level ordering generating
the $N=34$ shell closure.

If we change $R$  from $4.3\,$fm
to $4.5\,$fm the $1f$ and $2p$ levels
are pushed down but their energy difference increases, 
the $2p_{1/2}$ level lies above the $1f_{5/2}$, and therefore we do not have any more a shell closure at $N=34$ (see the right column of fig. \ref{fig:ws}).

In this work we present the results that we have obtained 
by investigating the emergence of this new magic number by using
a non-relativistic IPM approach which uses a Hartree-Fock (HF) plus
Bardeen-Cooper-Schrieffer (BCS) model for the ground states, and
a Quasi-particle Random Phase Approximation (QRPA) for the description
of the excited states. The sets of s.p. wave functions have been
generated by carrying out (HF) calculations with
density dependent finite-range interactions of Gogny type. We used these
results as input of a BCS calculation in order
to take into account the effects of the pairing. 
In refs. \cite{ang16a,ang16b,ang19} we have 
shown the good agreement between
the results of our HF+BCS approach with those of Hartree-Fock-Bogoliu\-bov
(HFB) calculations.

The HF s.p. wave functions
and the BCS output, in terms of changes of the HF s.p. energies and modifications 
of the occupation numbers of the HF s.p. levels, have been used
in QRPA calculations in order 
to obtain the excitation energies of the 2$^+$ states. 
A detailed description of the QRPA formalism is given in the work by De Donno et al. \cite{don17}.

We consistently use the same finite-range interaction in all the three 
steps of our calculations. Specifically, we have carried 
out the present investigation by using four 
different parameterizations of the density-dependent finite-range
Gogny interaction. These parameterizations are those known
in the literature as D1S \cite{ber91},  D1M \cite{gor09},
D1ST2a \cite{gra13} and  D1MTd \cite{co18a}. 
The parameters of the first two forces were selected, 
with a procedure described in detail by Chappert \cite{cha07t}, 
by fitting binding energies, root mean square (rms) charge 
radii and fission properties of a wide set of nuclei. 
In the present context it is relevant to remark that 
the strength of the spin-orbit term of the interaction was 
adjusted to reproduce the experimental 
energy splitting between the $1p_{1/2}$ and $1p_{3/2}$ 
neutron s.p. levels in $^{16}$O.

The other two forces, D1ST2a and  D1MTd 
have been constructed by adding to the D1S and
D1M interactions, respectively, 
one isospin dependent and one isospin independent 
tensor components. 
These tensor terms modify sensitively the splitting 
of the s.p. spin-orbit partner levels \cite{ots06},
but do not affect most of the quantities considered in the force fitting above mentioned.
The strengths of the spin-orbit 
and tensor terms of  the D1MTd force were chosen to properly describe
the excitation energies of the low-lying $0^-$ states in $^{16}$O and 
in $^{48}$Ca \cite{co18a}, while those of the D1ST2a 
were selected to reproduce the excitation energy of the $0^-$ states in $^{16}$O
and the energy splitting between the $1f$ s.p. states in $^{48}$Ca \cite{gra13}.

By using the approach described above, we investigated the region 
of the nuclear chart where changes in
the ordering of the $1f_{5/2}$ and $2p_{1/2}$ levels may occur. 
Specifically, we have calculated the evolution of the s.p. energies of 
these two levels for the sequence of $N=34$
isotones from $^{48}$Si to $^{66}$Ge.

All these nuclei have been considered to be spherical. 
While this permits a rather good description of the slightly deformed
$^{58}{\rm Cr}$, $^{60}{\rm Fe}$ and $^{62}{\rm Ni}$
nuclei, it turns out to be a rather poor approximation for the more heavily deformed 
$^{64}{\rm Zn}$ and $^{66}{\rm Ge}$ nuclei.
The agreement with the available experimental data is, however, rather
satisfactory. Our calculations slightly underbind the nuclei investigated but 
the largest relative difference with the experimental binding energy 
is of about 1.3\%. It is not a surprise that this is the value we found for
the two most deformed nuclei quoted above.

\begin{figure}[!b]
\centerline{\includegraphics[width=3.5in]{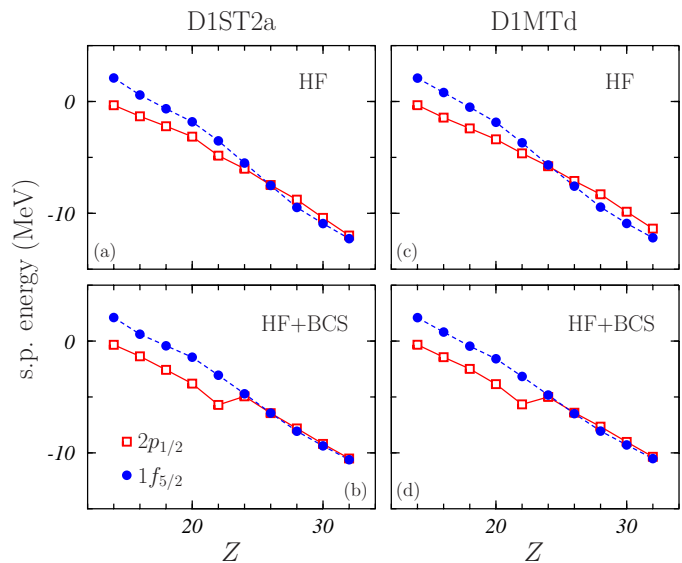}}
\vspace*{8pt}
\caption{Energies of the $1f_{5/2}$ (blue solid circles) and of the $2p_{1/2}$ (red open squares) calculated in HF (panels (a) and (c)) and in HF+BCS (panels (b) and (d)) with the D1ST2a (panels (a) and (b)) and the D1MTd (panels (c) and (d)) interactions for a sequence of $N=34$ isotones from the $^{48}{\rm Si}$ to the $^{66}{\rm Ge}$.\protect\label{fig:fp}}
\end{figure}

We show in the panels (a) and (c) of fig. \ref{fig:fp} the s.p. energies $\epsilon_k$ 
of the $1f_{5/2}$ and $2p_{1/2}$ levels, obtained in
our HF calculations by using the D1ST2a and D1MTd
interactions presented above,
for the aforementioned set of $N=34$ even-even isotones. 
For comparison, in the panels (b) and (d) of the same figure, 
we show the pairing corrected HF+BCS quasi-particle energies defined as
\begin{equation}
\epsilon^{\rm HF+BCS}_k\,=\,\left\{ 
\begin{array}{ll}
 \lambda \,-\, E^{\rm q}_k \, , & {\rm if}\,\, \epsilon_k < \lambda \, , \\
  & \\
 \lambda \,+\, E^{\rm q}_k \, , & {\rm if}\,\, \epsilon_k > \lambda \, , 
\end{array} 
\right.
\label{eq:spenergy}
\end{equation}
where $\lambda$ is the chemical potential obtained 
in the BCS calculation  
and $E^{\rm q}_k$ is the corresponding quasi-particle energy.
Similar results are obtained with the D1S and D1M interactions.

As seen in fig. \ref{fig:fp}, the various calculations
predict that in the nuclei with $Z < 26$ the s.p. energy of the 
$2p_{1/2}$ level (red open squares)
is lower than that of the $1f_{5/2}$ level (blue solid circles). 

This behavior is rather general, and is related to the increase 
of the radial dimensions of the neutron effective potential generated by HF 
calculations when the proton number increases. We tested this fact by carrying out  
HF calculations with the Skyrme SkI interaction for a fixed number of neutrons
and by increasing the proton numbers. We observed a growth of the radial 
dimensions of the effective potential and, for the s.p. energies, we obtained 
the same effect shown in fig. \ref{fig:ws} for a simple Woods-Saxon potential. 
We can state that the observed trend is an overall effect of the nuclear mean 
field with small differences among the different effective nuclear interactions considered.

These results indicate that the best nuclei to 
identify the emergence of the new magic number 34 are those 
with the smallest proton number $Z$ where the energy difference
between the $1f_{5/2}$ and the $2p_{1/2}$ states is accentuated. This 
makes the $^{48}$Si nucleus a good
testing ground for these studies, as pointed out by Li et al. \cite{li18}.

We investigated the 
peculiar properties of this nucleus by considering how 
pairing effects evolves in the silicon isotope chain.
We considered only even-even nuclei from $^{32}$Si
up to $^{48}$Si \/ and assumed a spherical shape for all these isotopes.
While the deformations of  $^{40}$Si and $^{44}$Si are relatively
small, those of $^{42}$Si and $^{44}$Si are instead remarkable 
\cite{cea}. 

\begin{table}[!t] 
%\begin{center} 
{\begin{tabular}{@{}cccccccc@{}} 
\toprule
 $N$ & $A$ && D1S & D1ST2a & D1M & D1MTd & exp \\ 
 \colrule
 18 & 32 &&  -268.8 & -265.6 & -265.1& -264.8 & -271.4 \\
 20 & 34 && -282.3 & -283.2 & -279.9 & -282.3 & -283.4 \\
 22 & 36 && -288.9 & -287.3  & -285.9 &-286.7 &-292.1  \\
 24 & 38 && -296.1  &-292.8 &-292.8 &-292.5  &-299.9 \\
 26 & 40 && -302.4 & -297.6  &-299.2 &-297.6 &-306.5 \\
 28 & 42 && -307.7 & -302.2  &-304.6 &-302.0 &-311.6 \\
 30 & 44 && -309.8 & -304.4 & -306.7 &-304.4 &-315.7 \\
 32 & 46 && -311.5 & -305.7 & -308.7  &-306.5  & \\
 34 & 48 && -311.2 & -305.7 & -308.5 &-306.7 & \\
\botrule
\end{tabular}
}
%\end{center} 
\caption{Total HF+BCS binding energies, in MeV, 
of the Si isotopes under investigation.
The experimental values are taken from the Brookhaven National Lab compilation \cite{bnlw}.
\label{tab:drip} }
\end{table} 

We show in table \ref{tab:drip} the total HF+BCS binding energies of the 
Si isotopes under investigation obtained with the four interactions
we have considered. In the cases of D1S and D1M interactions the 
binding energy of $^{48}$Si is slightly smaller than that of $^{46}$Si
nucleus. This indicates that, for these two forces, $^{48}$Si is beyond
the two-neutron drip line. The situation is reversed for the 
D1MTd interaction, while for the D1ST2a force both nuclei have the same binding energy.

By changing of 10\% the numerical input parameters, such
as integration range and number of integration points in both radial 
and momentum space, we estimated the numerical uncertainty of our calculations. 
In the case of the binding energies it resulted of about 0.3\% and 
then the small differences obtained between the total binding 
energies are within our numerical accuracy. 

The results of table \ref{tab:drip} 
indicate that, in our model, the $^{48}$Si nucleus 
is at the edge of the two-neutron emission drip line. 
The HFB results of the Amedee database \cite{cea}, obtained
with the Gogny D1S force, places $^{48}$Si beyond the drip line.
On the other hand, the results presented in the supplementary
information of the work by Erler et al. \cite{erl12}, obtained
with various parameterizations of the Skyrme interaction, indicate
that the $^{48}$Si nucleus is positioned before the drip line 
for all the forces considered but for SLy4. The results of 
HFB calculations we carried out with the HFBRAD code  \cite{ben05}
indicate that also the SIII force predicts a $^{48}$Si nucleus beyond the
drip line. The recent compilation of relativistic HFB results of Xia et al. \cite{xia18}
finds $^{48}$Si before the two-neutron drip line. 
Then, the position of the drip line in Si isotopes is an 
open problem from the theoretical point of view.

We continued our investigation by considering henceforth only
the two interactions containing tensor terms, D1ST2a and D1MTd,
which generate a $^{48}$Si stable against the
two-nucleon emission.

\begin{table}[!b] 
{\begin{tabular}{@{}cc c ccc c ccc@{}} 
\toprule
             &   & \phantom{p} & \multicolumn{3}{c}{D1ST2a} &~~~~~~& \multicolumn{3}{c}{D1MTd} \\ \cline{4-6} \cline{8-10}
\rule{0cm}{0.4cm} $N$ & $A$ && $\Delta N^2 $& $-B_{\rm p} / A$ (MeV) & $\Delta r$ (\%) & \phantom{pp}
                   & $\Delta N^2 $& $-B_{\rm p} / A$ (MeV) & $\Delta r$ (\%) \\
\colrule
18 & 32 & & 2.66                    & 0.11                    & $2\cdot 10^{-3}$  & &  2.62                    & 0.10                    & $2\cdot 10^{-3}$\\
20 & 34 & & $4\cdot 10^{-4}$ & $3\cdot 10^{-4}$ & $1\cdot 10^{-6}$ & & $1\cdot 10^{-4}$ & $2\cdot 10^{-4}$ & $7\cdot 10^{-7}$\\
22 & 36 & & 3.24                    & 0.09                    & $4\cdot 10^{-3}$  & &  3.22                   & 0.08                     & $3\cdot 10^{-3}$\\
24 & 38 & & 4.36                    & 0.12                    & $6\cdot 10^{-3}$  & &  4.29                   & 0.10                     & $4\cdot 10^{-3}$\\
26 & 40 & & 3.41                    & 0.11                    & $5\cdot 10^{-3}$  & &  3.32                   & 0.08                     & $4\cdot 10^{-3}$\\
28 & 42 & & 0.25                    & 0.09                    & $2\cdot 10^{-3}$  & &  0.04                   & 0.04                     & $3\cdot 10^{-4}$\\
30 & 44 & & 2.39                    & 0.12                    & $4\cdot 10^{-3}$  & &  2.24                   & 0.07                     & $2\cdot 10^{-3}$\\
32 & 46 & & $6\cdot 10^{-4}$ & 0.10                    & $8\cdot 10^{-6}$  & & $8\cdot 10^{-4}$ & 0.07                     & $2\cdot 10^{-6}$\\
34 & 48 & & $7\cdot 10^{-4}$ & 0.10                    & $2\cdot 10^{-5}$  & & $5\cdot 10^{-5}$ & 0.07                     & $7\cdot 10^{-7}$\\
\botrule
\end{tabular}
\label{tab:silicon} }
\caption{Some quantities related to pairing effects calculated for the silicon isotope chain. Each nucleus  
is here identified by the neutron and mass numbers, $N$ and $A$, respectively. The quantity
$\Delta N^2$ indicates the fluctuation of the number of particles. The pairing energy, $B_{\rm p} / A$, is the contribution of the
pairing to the total nuclear binding energy per nucleon. Finally, $\Delta r$, defined in eq. (\ref{eq:Deltar}), gives the relative difference between the neutron rms radii obtained in HF+BCS and HF calculations.}
\end{table} 

One of the key features of the magic numbers is the 
absence, or the strong reduction, of pairing effects.
For the set of silicon isotopes that we investigated,
we collect in table \ref{tab:silicon} the values of some 
quantities indicative of the relevance of the pairing.
With $\Delta N^2$ we indicate the fluctuation of the particle number
in BCS calculations. The rows labelled as $B_{\rm p}/A$ show
the contribution of the pairing to the nuclear binding energy 
per nucleon, expressed in MeV. Finally, 
\begin{equation}
\Delta r \,= \, \displaystyle \frac{|r^{\rm HF+BCS}\, -\, r^{\rm HF}|}{r^{\rm HF}}
\label{eq:Deltar}
\end{equation}
gives the relative difference between the rms radii calculated 
in HF+BCS and in HF calculations. 

All the three quantities under investigation show
analogous behaviors for both interactions. 
The smallest values, indicating
small pairing effects, appear for $N=20$, corresponding 
to the closure of the $1d_{3/2}$ level, and
for $N=28$, due to the   
complete occupancy of the $1f_{7/2}$ s.p. state. 
The values of the three quantities increase for
the nucleus $^{44}{\rm Si}$ where
the $2p_{3/2}$ level is only half filled. 
The full occupancy of this level strongly 
reduces the pairing effects for $N=32$ 
and indicates a shell closure that has been already investigated in 
Ca and Ti isotopes \cite{jan02,lid04a,for04,for05a,hon05,wie13,ste13,lid04b}.
 A similar situation occurs for $N=34$ where the $2p_{1/2}$ is completely filled 
and the extremely low values of $\Delta N^2$ and $\Delta r$ point to it as a magic number. 

The isotopic evolution of the first $2^+$ excited state may also indicate a shell closure
 \cite{gad08a,sor08}. We have calculated the $2^+$ excited states of various Si isotopes
by using the QRPA.  In this approach, the nuclear excited states are described as
linear combinations of one-particle one-hole, one-particle one-particle, or one-hole one-hole configurations. 
More complicated configurations, such as $n$-particle $n$-hole or $n$-particle $n$-particle configurations, are neglected.

The QRPA results are very sensitive to the dimensions
of the s.p. configuration space. 
We have selected its size to avoid imaginary energy solutions and
to reproduce at best the behavior of  the known experimental energies,
i.e. those of the $^{40}$Si  and $^{42}$Si nuclei.

Following these criteria, we selected all the particle-hole configurations 
with excitation energy differences smaller than $30\,$MeV and the 
particle-particle pairs with maximum excitation energy of $10\,$MeV.
In addition, we did not consider those
particle-particle pairs where the product of their occupation 
numbers is smaller than $10^{-4}$. 
This last restriction reduces noticeably the particle-particle
configuration space but it changes the energies of the first 
$2^+$ excited states by less than 1\%.

\begin{table}[!b] 
{\begin{tabular}{@{}cc c cc c cc c c@{}} 
\toprule
             &   &~~~& \multicolumn{2}{c}{D1ST2a} &~~~& \multicolumn{2}{c}{D1MTd} &~~~& experiment\\ 
             \cline{4-5} \cline{7-8} \cline{10-10}
 & && $\omega$ & $B(E2)$  && $\omega$ &  $B(E2)$   && $\omega$ \\
 $N$ & $A$ && (MeV)& (e$^2$ fm$^4$) && (MeV)&  (e$^2$ fm$^4$) && (MeV) \\ \colrule
26 & 40  && 1.45 &  81.9  &&  1.49 &  84.7  && 0.986 \\ 
28 & 42  && 1.24 &  37.5   &&  1.28 &   2.4 && 0.77 \\  
30 & 44  && 0.45 &  186.2   &&   0.52 &  101.5  &&  \\ 
32 & 46  && 1.24  & 70.4  &&  1.10 &   65.3 && \\  
34 & 48  && 1.61 &  30.1   &&   1.61 &  24.7 && \\
\botrule
\end{tabular}
\label{tab:s2p} }
\caption{QRPA results for the excitation of the first $2^+$ state
in some Si isotopes calculated by using the D1ST2a and
the D1MTd interactions. We have indicated with $\omega$ the excitation energy. The experimental values quoted for the
$^{40}$Si and $^{42}$Si isotopes are from Campbell {et~al.} \cite{campbell2006} and Bastin {et~al.} \cite{bastin2007}, respectively.
}
\end{table} 

We show in table \ref{tab:s2p} the excitation energies, $\omega$, 
and the $B(E2)$ values
of the first $2^+$ excited state of the even-even Si isotopes 
from $A=40$ up to $A=48$, obtained with the D1ST2a and D1MTd interactions,
as well as the two known experimental excitation energies. The energy difference
between these two energies is well reproduced by our calculations, 
even though the values that we have obtained are shifted by $\sim 0.5\,$MeV. 
This may be due to the fact that both $^{40}$Si  and $^{42}$Si are deformed 
nuclei \cite{cea}.

The largest excitation energy of the $2^+$ states shown in table \ref{tab:s2p} 
is that of the $N=34$ 
isotope. This occurs for both the interactions considered and it can be
considered as an additional hint of the $N=34$ shell closure.

In conclusion, we have investigated the 
emergence of a new magic number at $N=34$. 
For nuclei far from the stability valley with low $Z$, 
the corresponding shell closure occurs because the energy 
of the s.p. $2p_{1/2}$ state is 
smaller than that of the $1f_{5/2}$ level up to $Z \sim 24$ 
where the ordering of these two s.p. states is reversed. 
Even more, the lower the $Z$ value is, the larger becomes the energy gap, 
thus favoring the appearance of the new magic number.
This effect is related to the radial dimension of the neutron 
effective potential that increases with $Z$ along the $N=34$ isotonic chain.

Our non-relativistic calculations confirm the findings 
of Li et al. \cite{li18} indicating the $^{48}$Si nucleus as a good candidate 
to identify the shell closure at $N=34$. However, it is worth pointing out that
this nucleus is very close to the two-neutron emission drip line. 
For example, our results for the D1S and D1M
forces indicate that the $^{48}$Si is beyond this line, while 
the two interactions with tensor terms, D1ST2a 
and D1MTd, situate it before the drip line.

In any case, it must to be noted that, although the results seem discrepant,
the differences between them are within the 
limit of the numerical accuracy of our calculations. 
In these circumstances we believe that to elucidate experimentally
the positioning of the drip line in Si isotopes
and the eventual existence of the $^{48}$Si nucleus would be a challenging 
investigation that could shed light on the behavior of the shell evolution
far from the stability valley, and set stringent limits to the nuclear
theories.

\section*{Acknowledgments}

This work has been partially supported by  
the Junta de Andaluc\'{\i}a (FQM387), the Spanish Ministerio de 
Econo\-m\'{\i}a y Competitividad (FPA2015-67694-P) and the European 
Regional Development Fund (ERDF).

\end{document}